# Efficient Spin Transfer in $WTe_2/Fe_3GeTe_2$ van der Waals Heterostructure Enabled by Direct Interlayer p-Orbital Hybridization

H. L. Ning[1,#], M. Q. Dong[2,#], X. Zhang[3,†], J. S. Huang[1], B. Liu[2], Zhi-Xin Guo[1,*], Y. Dong[4,‡]

[1]*State Key Laboratory for Mechanical Behavior of Materials, School of Materials Science and Engineering, Xi'an Jiaotong University, Xi'an, Shaanxi, 710049, China.*

[2]*Key Laboratory of Computational Physical Sciences (Ministry of Education), Institute of Computational Physical Sciences, State Key Laboratory of Surface Physics, and Department of Physics, Fudan University, Shanghai 200433, China*

[3]*Shaanxi Key Laboratory of Surface Engineering and Remanufacturing College of Mechanical and Materials Engineering Xi'an University, Xi'an 710065, China.*

[4]*State Key Laboratory of Oral & Maxillofacial Reconstruction and Regeneration, National Clinical Research Center for Oral Diseases, Shaanxi Key Laboratory of Stomatology, Department of Prosthodontics, School of Stomatology, The Fourth Military Medical University, Xi'an, Shaanxi 710032, China.*

[#]These authors contributed equally to this work
[†]zhangxian1@outlook.com
*zxguo08@xjtu.edu.cn
[‡]dongyanfmmu@qq.com

## Abstract

Recent experiments have demonstrated efficient spin transfer across layers in the van der Waals heterostructure composed of $WTe_2$ and $Fe_3GeTe_2$, signaling a potential breakthrough in developing all-van der Waals spin-orbit torque devices. However, the reasons behind the unusually high interlayer spin transparency observed, despite the weak van der Waals interactions between layers, remain unclear. In this study, we employ density functional theory and the non-equilibrium Green's function method to explore this phenomenon. We find that the efficient cross-layer spin transfer arises from direct hybridization of p-orbitals between tellurium atoms at the interface. This interlayer orbital hybridization lowers the electronic potential barrier and significantly modifies the spin-polarized electronic structure of $Fe_3GeTe_2$. Consequently, an effective channel for spin-polarized transport is established between $WTe_2$ and $Fe_3GeTe_2$, leading to high interlayer spin transparency. Combining this enhanced spin transparency with the large spin Hall angle of $WTe_2$ explains the high spin-orbit torque efficiency observed experimentally. Furthermore, we predict that applying a gate voltage can further increase this efficiency. Our findings offer a pathway for designing high-performance, all-van der Waals spin-orbit torque devices.

## Ⅰ. INTRODUCTION

Spintronics, a promising next-generation information technology, relies on the effective generation and injection of spin currents. Spin-orbit torque (SOT), induced by out-of-plane spin-polarized currents generated in materials with a high spin Hall effect (SHE) and injected into a ferromagnetic (FM) layer, has been successfully employed to efficiently manipulate magnetization, enabling the development of commercial SOT-MRAM devices [1–3]. Significant efforts have been dedicated to exploring novel materials for high-performance SOT devices [4–8]. A critical challenge in achieving efficient SOT-based spintronic devices is maximizing the SOT efficiency, defined as $\xi = \theta_{SH} \cdot T_{int}$ [9], where $\theta_{SH} = |J_s|/|J_c|$ represents the spin Hall angle, quantifying the conversion efficiency of charge current density ($J_c$) into spin current density ($J_s$). $T_{int}$ is the spin transparency across the interface between the high-SHE material and the FM layer.

To date, various high spin-orbit coupling (SOC) materials, including heavy metals (e.g., W, Ta, Pt) [10–13], topological insulators (TIs) [14–16], and more recently, topological semimetals (TSMs) [17–20], have been extensively studied to optimize their $\theta_{SH}$. However, investigations of $T_{int}$ remain scarce. This is primarily because FM materials commonly used in SOT devices, such as Co, Fe, and their alloys (e.g., CoFeB), are metallic and exhibit strong interface interactions with SOC materials, which are generally assumed to result in efficient spin transparency (i.e., large $T_{int}$).

In parallel, rapid advancements in van der Waals (vdW) materials and their heterostructures offer new opportunities for enhancing spintronic functionalities. Recently, various two-dimensional (2D) ferromagnetic (FM) vdW materials have been discovered [21,22], including $CrX_3$ (X = I, Cl, and Br) [23-25], $Cr_2Ge_2Te_6$ [26-28], and $Fe_3GeTe_2$ (FGT) [29-31]. Among these, 2D FGT has garnered significant attention due to its relatively high Curie temperature and perpendicular magnetic anisotropy (PMA) [32,33], characteristics that make FGT a promising candidate for applications in 2D spintronic devices.

Very recently, a remarkable breakthrough in all-vdW SOT devices has been reported based on the $WTe_2$/FGT heterostructure [33]. This all-vdW SOT device exhibited an exceptionally high SOT efficiency ($\xi = 4.6$), which is significantly greater than that of traditional SOT devices, such as Pt/CoFeB ($\xi = 0.093$) and Ta/CoFeB ($\xi =$ 0.05-0.08) , as well as recently developed “half-vdW” SOT devices, including $WTe_2$/Py ($\xi$ = 0.09-0.51), Pt/FGT ($\xi$ = 0.12-0.14), and $WTe_x$/CoFeB ($\xi$ = 0.43) [34–38].This discovery is particularly striking because the interface interaction between the FM and SOC materials is weakest in all-vdW SOT devices (vdW bonding) compared to traditional and “half-vdW” devices (chemical bonding). Typically, weaker interface interactions correlate with a smaller $T_{int}$ and consequently lower SOT efficiency. Therefore, understanding the underlying mechanism responsible for the unusually high SOT efficiency in the $WTe_2$/FGT heterostructure is crucial and would provide significant guidance for the further development of high-performance all-vdW SOT devices.

In this work, we employ density functional theory (DFT) and non-equilibrium Green's function (NEGF) calculations to investigate the interfacial interaction and cross-plane electronic transport properties of the $WTe_2$/FGT heterostructure. Despite the weak vdW interlayer interaction, our results reveal significant p-orbital hybridization between Te atoms in $WTe_2$ and FGT at the interface. This pronounced interlayer orbital hybridization establishes an effective spin-polarized transport channel between the two layers. Consequently, the heterostructure exhibits an unusually large cross-plane $T_{int}$ for spin-up electrons. Combined with the large $\theta_{SH}$ in $WTe_2$, this leads to the observed high SOT efficiency.

## Ⅱ. METHOD

Electronic structure calculations of 1T'-$WTe_2$/FGT vdW heterostructure are performed within the framework of ab initio DFT using the projector-augmented wave (PAW) potentials, as implemented in Vienna Ab initio Simulation Package (VASP) [39-41]. The Perdew-Burke-Ernzerhof (PBE) exchange-correlation function is taken into account with generalized gradient approximation (GGA) to determine the exchange-correlation coupling [42]. The semiempirical dispersion-corrected DFT scheme of Grimme (DFT-D3 approximation) is applied to evaluate the weak interlayer vdW interaction [43]. The vacuum region is set to be about 15 Å in the Z-direction to address the image of charge transfer and eliminate the interaction between the periodic boundary. A plane-wave energy cutoff of 500 eV ensures total energy convergence, with an energy convergence threshold of $10^{-6}$ eV for both monolayers and the heterostructure. The reciprocal space is sampled by a fine grid of 14×14×1, 16×10×1 and 9×9×1 Monkhorst-Pack k points in the Brillouin zone (BZ) for geometrical optimization for monolayer FGT, monolayer $WTe_2$ and their heterostructure, respectively [44]. The GGA + U method accounts for the localized 3d orbitals, with $U_{eff}$ = 4 eV selected for Fe 3d orbitals based on linear response calculations and consistency with previous studies [45]. Note that similar U values (3.5-4.3 eV) have also been adopted in other works on FGT-based heterostructures, yielding consistent and physically meaningful results [45-49]. Structural relaxations are carried out until the force on each atom is less than 0.01 eV/Å. In the DFT calculations, lattice constants of a = 3.49 Å and b = 6.28 Å are used for $WTe_2$ [50], while a = b = 3.99 Å are used for FGT [51].

The quantum transport properties are calculated based on DFT combined with the Keldysh NEGF formalism, as implemented in the Atomistix ToolKit (ATK) 2019 package [52]. In the NEGF-DFT transport simulation, the physical quantities are expanded by a linear combination of atomic orbital (LCAO) basis set [53,54]. PseudoDojo pseudopotential and GGA in the form of PBE function are utilized to represent the exchange and correlation interactions [55]. The spin-resolved conductance is obtained by the Landauer-Buttiker formula:

$$G_\sigma = \frac{e^2}{h} \sum_{k_{||}} T_\sigma\left(k_{||}, E_F\right) \quad (1)$$

where $T_\sigma(k_{||}, E_F)$ is the transmission coefficient with spin σ at the transverse Bloch wave vector $k_{||} = (k_x, k_y)$ and the Fermi level $E_F$, e is the electron charge and h is the Planck constant. 11 × 11 × 1 Monkhorst-Pack k-point mesh is used. The real-space mesh cut-off is set to be 120 Hartree. The numerical tolerance of the Hamiltonian matrix of self-consistency is restricted to $10^{-4}$.

## Ⅲ. RESULT AND DISCUSSION

We first present the atomic and electronic properties of monolayer $WTe_2$ and FGT. The monolayer $WTe_2$, naturally obtained via mechanical stripping, adopts a 1T' phase with spatial inversion symmetry. In this phase, spontaneous Peierls distortions cause W atoms to deviate from their ideal positions, forming a distorted octahedral structure [56]. As a result, 1T'-$WTe_2$ exhibits mirror symmetry only for the bc plane, but not for the ac plane [Fig. 1(a)]. The projected band structure of monolayer $WTe_2$ is shown in Fig. 1(b) and Fig. S1(a). $WTe_2$ displays metallic characteristics, with energy bands near the $E_F$ predominantly contributed by W atoms. Additionally, energy bands associated with Te atoms, characterized by strong p-d hybridization with W atoms, also appear around $E_F$, reflecting the shallow energy level of p orbitals in $WTe_2$.

For the monolayer FGT system (Fig. 1(c)), we determine the magnetic ground state by comparing the total energies of four typical magnetic configurations (Fig. 2). As shown in Table 1, the ferromagnetic (FM) state is found to be energetically more favorable than the three antiferromagnetic (AFM) states, in agreement with experimental observations that identify FGT as a FM material at the ground state [57-60]. Furthermore, our calculations show that the easy axis of magnetization for freestanding FGT lies along the out-of-plane (z) direction, corresponding to a perpendicular magnetic anisotropy (PMA) of 1.0 meV per Fe atom. Figs. 1(d) additionally shows the projected band structure of monolayer FGT, which exhibits metallic behavior with most energy bands crossing $E_F$ contributed by the spin-down electrons of Fe atoms. Nonetheless, several energy bands from the spin-up electrons of Fe and Te atoms also cross $E_F$. The presence of Te-characteristic energy bands near $E_F$ (Fig. S1(b)) indicates the high activity of p orbitals in Te atoms within FGT. Considering the shallow energy level of p orbitals in Te atoms of both $WTe_2$ and FGT, significant p-p orbital hybridization is expected between Te atoms at the interface of the heterostructure. This hybridization is stronger than the interfacial orbital interactions commonly observed in other van der Waals heterostructures.

Next, we discuss the $WTe_2$/FGT heterostructure. The in-plane lattice constants of freestanding $WTe_2$ and FGT are a = 3.48 Å, b = 6.25 Å and a = b = 3.99 Å, respectively. A $2 \times 2/\sqrt{3}$ super-periodicity of $WTe_2$ is commensurate with the $\sqrt{3} \times \sqrt{3}$ FGT lattice, resulting in lattice mismatches of 0.7% and 4.2% along the a- and b-directions, respectively. As shown in Fig. S2, we consider three typical stacking configurations for the $WTe_2$/FGT heterostructure: hollow, bridge, and top configurations. Total-energy calculations reveal that the hollow configuration (Fig. S2(a)) is the most stable, being 2.21 meV and 140.02 meV more stable than the bridge and top configurations, respectively. Consequently, all subsequent studies are based on the hollow configuration. The minimum distance between the nearest interlayer Te atoms of $WTe_2$ and FGT in this configuration is 3.73 Å. Given that the atomic radius of Te is 1.37 Å, the sum of their atomic radii is significantly smaller than the interlayer spacing. This indicates that the interlayer interaction between $WTe_2$ and FGT is of the vdW type. It is noteworthy that the collinear state is more stable than non-collinear states (see Section SIII of the Supplemental Material [61] for details), and the magnetic ground state of the $WTe_2$/FGT heterostructure remains ferromagnetic, as confirmed by total energy comparisons (see Table 1), in agreement with experimental observations [33]. A sizable PMA of approximately 0.58 meV per Fe atom is also obtained. This robust out-of-plane magnetic anisotropy highlights the potential of the $WTe_2$/FGT heterostructure for technological applications.

To confirm the energetic stability of the $WTe_2$/FGT heterostructure, we calculate the binding energy ($E_b$) between $WTe_2$ and FGT. The binding energy is defined as $E_b = (E_{tot} - E_W - E_F)/S$, where $E_{tot}$, $E_W$ and $E_F$ represent the total energy of the $WTe_2$/FGT heterostructure, monolayer $WTe_2$ and monolayer FGT, respectively. S denotes the cross-sectional area of the system. For the hollow configuration, we find $E_b$ = -190 meV/Å$^2$, indicating that the heterostructure is energetically stable under vdW interlayer interactions. Notably, the binding energy of the $WTe_2$/FGT heterostructure is significantly larger than that of typical vdW materials, such as bilayer graphene (-19 meV/Å$^2$) [70], h-BN (-19 meV/Å$^2$) [71], and $MoS_2$ (-40 meV/Å$^2$) [71]. This suggests that the $WTe_2$/FGT heterostructure may exhibit more efficient interlayer electronic transport properties compared to conventional vdW materials. Note that the binding energies of chemical bonding interactions are typically in the eV range, which far exceeds the scale of vdW interactions. Therefore, the interfacial interaction in our system remains within the vdW category. In addition, details of the phonon spectrum calculations are provided in Section SIV of the Supplemental Material [61]. The results confirm that the $WTe_2$/FGT heterostructure is dynamically stable.

We further investigated the effect of interfacial interactions on the magnetic stability of FGT by employing a Heisenberg model, with the Hamiltonian defined as [65,72]:

$$H = E_0 + \sum_{ij} J_z \boldsymbol{S}_i \boldsymbol{S}_j + \sum_{iK} J_{top} \boldsymbol{S}_i \boldsymbol{S}_K + \sum_{jK} J_{bot} \boldsymbol{S}_j \boldsymbol{S}_K + A S_i^z S_i^z. \quad (2)$$

Here $\boldsymbol{S}_i$ denotes the spin vector of each magnetic atom, and $S_i^z$ is the z component. A represents the magnetic anisotropy energy (MAE), for which we adopt the previously calculated PMA values of 0.58 and 1.0 meV for FGT and $WTe_2$/FGT, respectively. The exchange parameters $J_z$, $J_{top}$ and $J_{bot}$ correspond to interactions between two $Fe^{3+}$ atoms in the bottom and top layers, between $Fe^{2+}$ atoms in the middle layer and $Fe^{3+}$ atoms in the top layer, and between $Fe^{2+}$ and $Fe^{3+}$ atoms in the bottom layer, respectively (see Fig. 2). Note that $J_{top} = J_{bot}$ in the freestanding FGT, while $J_{top} \neq J_{bot}$ in the $WTe_2$/FGT heterostructure due to the interface interaction.

Hence, the Heisenberg spin Hamiltonian for the four magnetic configurations shown in Fig. 2 can be expressed as:

$$E_{FM} = E_0 - 4J_z|S_{Fe1}|^2 - 12J_{top}|S_{Fe1}||S_{Fe2}| - 12J_{bot}|S_{Fe1}||S_{Fe2}| + A|S|^2, \quad (3)$$

$$E_{AFM1} = E_0 - 4J_z|S_{Fe1}|^2 + 12J_{top}|S_{Fe1}||S_{Fe2}| + 12J_{bot}|S_{Fe1}||S_{Fe2}| + A|S|^2, \quad (4)$$

$$E_{AFM2} = E_0 + 4J_z|S_{Fe1}|^2 - 12J_{top}|S_{Fe1}||S_{Fe2}| + 12J_{bot}|S_{Fe1}||S_{Fe2}| + A|S|^2, \quad (5)$$

$$E_{AFM3} = E_0 + 4J_z|S_{Fe1}|^2 + 12J_{top}|S_{Fe1}||S_{Fe2}| - 12J_{bot}|S_{Fe1}||S_{Fe2}| + A|S|^2, \quad (6)$$

where $E_{FM}$, $E_{AFM1}$, $E_{AFM2}$ and $E_{AFM3}$ denote the total energies of the FM ground state (Fig. 2(a)) and the three AFM configurations (Figs. 2(b-d)), respectively. With $|S_{Fe1}| = \frac{3}{2}$ and $|S_{Fe2}| = \frac{2.5}{2}$, one can calculate $J_z$, $J_{top}$ and $J_{bot}$ in use of $E_{FM}$, $E_{AFM1}$, $E_{AFM2}$ and $E_{AFM3}$ based on Eqs. (3)–(6) [73]. As a result, for freestanding FGT, the exchange parameters are $J_z$=6.68 meV and $J_{top}$=$J_{bot}$=14 meV. In contrast, for FGT in the heterostructure, we obtain $J_z$=57.16 meV, $J_{top} = 8.12$ meV and $J_{bot}$=5.32 meV. These results indicate that the interlayer interaction induced by $WTe_2$ significantly enhances the exchange coupling between $Fe^{3+}$ ions, while it suppresses the exchange interaction between $Fe^{2+}$ and $Fe^{3+}$ ions.

Based on this model, we further estimated the Curie temperatures ($T_C$) of both monolayer FGT and $WTe_2$/FGT heterostructure using Monte Carlo simulations [74]. Figure 3 presents the temperature dependence of the specific heat ($C_v$) for both systems, with the peak positions corresponding to $T_C$. The calculated $T_C$ values are approximately 110 K for freestanding FGT (Fig. 3(a)) and 123 K for FGT within the heterostructure (Fig. 3(b)). This enhancement in $T_C$ indicates that interfacial coupling with $WTe_2$ improves the thermal stability of ferromagnetism in FGT, which is favorable for potential applications.

We now turn to the electronic properties of the $WTe_2$/FGT heterostructure. Figs. 4(c) and 4(d) show the projected spin-polarized band structures of $WTe_2$ and FGT, respectively. Notably, the band structures of both components within the heterostructure exhibit significant deviations from those of their freestanding monolayer counterparts. This result indicates that the weak vdW interlayer interaction induces a substantial modification of the band structure in the $WTe_2$/FGT heterostructure. Additionally, we calculate the electron transfer between $WTe_2$ and FGT and find that only a small amount of charge (0.047 e per unit cell) transfers from FGT to $WTe_2$. This observation suggests that the significant band structure variation in the heterostructure is not driven by charge transfer. Instead, it is most likely attributed to orbital hybridization between interlayer Te atoms.

The significant orbital hybridization is further verified by the projected energy bands for the interlayer Te atoms. As shown in Fig. 5(a), there is a pronounced overlap of energy bands contributed by Te atoms of $WTe_2$ and FGT at interface for the spin-up electrons near $E_F$, as highlighted by the red circles. This observation is further supported by the calculated partial density of states (PDOS) of the interfacial Te atoms. As shown in Fig. 4(b), a pronounced spin-up PDOS is observed. Notably, there are clear overlaps between the PDOS peaks of $WTe_2$ and FGT at energy levels near -0.8 eV, 0.7 eV, and 0.5 eV, which align well with the features in the projected band structure shown in Fig. 5(c). This result aligns with our earlier discussion, confirming the strong p-p orbital hybridization between the Te atoms in $WTe_2$ and FGT at the interface for the spin. In contrast, for spin-down electrons, the energy bands from Te atoms of $WTe_2$ and FGT exhibit minimal overlap (Fig. 5(b)). This finding is also supported by the PDOS (Fig. 4(b)), where the peaks from $WTe_2$ and FGT rarely align at the same energy levels.

To further resolve the contributions of the $p_x$, $p_y$ and $p_z$ orbitals to the interfacial hybridization, we additionally calculated the orbital-resolved band structures weighted by the product $W_{pm1} \times W_{pm2}$, where $W_{pm1}$ and $W_{pm2}$ denote the projection weights of the $p_m$ (m=x, y, z) orbitals of the interfacial Te atoms in FGT and $WTe_2$, respectively. As shown in Fig. 5(c), a pronounced $p_z$-$p_z$ hybridization is observed at the interface for spin-up electrons. In contrast, for spin-down electrons, no significant interfacial hybridization of the p-orbitals is found near the Fermi level, as illustrated in Fig.5 (d). The calculated wave functions for the Te atoms, shown in Figs. 6(a) and 6(b), further illustrate this difference. For spin-up electrons, there is significant wave function overlap (primarily from $p_z$-orbitals) between interlayer Te atoms, whereas the overlap is negligible for spin-down electrons. This result is consistent with the PDOS analysis, indicating that the spin-up electronic states predominantly contribute to the interfacial electronic conductance. Moreover, the differential charge density (Fig. 6(c)) reveals a remarkable charge density redistribution in FGT, despite the limited interlayer charge transfer (0.047 e). This redistribution explains the substantial modulation of FGT's band structure in the $WTe_2$/FGT heterostructure.

Finally, we discuss the electronic transport properties of the $WTe_2$/FGT heterostructure. As shown in Fig. 6(d), the electrostatic potential barrier between $WTe_2$ and FGT in the heterostructure is 10.77 eV, indicating a Schottky-type interface contact. However, this barrier is significantly smaller than that of typical 2D vdW heterostructures composed of semiconductors and metals [64,75,76], which is attributed to the strong p-p orbital hybridization between interlayer Te atoms. A smaller potential barrier generally implies a higher electronic conductance. To further investigate the interfacial spin transport properties, we construct a 2D heterostructure device model, as illustrated in Fig. 7(a). The calculated spin-polarized transmission is shown in Fig. 7(b), where the transmission of the spin-up state is more than an order of magnitude larger than that of the spin-down state near $E_F$. This result is consistent with the analysis of spin-polarized electronic structures discussed above. This finding demonstrates that the $WTe_2$/FGT interface acts as an efficient spin filter. The spin current generated in $WTe_2$ via the SOT effect can be efficiently transferred into FGT, highlighting the potential of this heterostructure for spintronic applications.

Since the spin transfer efficiency is determined by both the spin Hall angle and spin transparency ($\xi = \theta_{SH} \cdot T_{int}$), with $\theta_{SH}$ primarily dependent on the intrinsic property of $WTe_2$, we mainly investigate the spin transfer efficiency of $WTe_2$/FGT heterostructure by examining $T_{int}$ between $WTe_2$ and FGT. The electronic conductivity of the heterostructure is found to be $2.74 \times 10^4$ S/m at the $E_F$. For comparison, we also examine the spin-transfer efficiency at the $WTe_2$-$WTe_2$ and FGT-FGT interfaces. The electronic conductance is calculated using the same device model, replacing the $WTe_2$/FGT heterostructure with $WTe_2$/$WTe_2$ and FGT/FGT, respectively. As shown in Fig. 7(c), the spin-up electronic conductance of the $WTe_2$/FGT heterostructure is obviously higher than that of $WTe_2$/$WTe_2$ around the $E_F$, demonstrating the excellent spin-transfer properties of the $WTe_2$/FGT interface. It is worth noting that the electronic conductance of both $WTe_2$/FGT and $WTe_2$/$WTe_2$ is lower than that of FGT/FGT, as FGT is a good metal with high electronic conductance. Nonetheless, the electronic conductance of $WTe_2$/FGT can be substantially enhanced by tuning the energy level (chemical potential). For instance, at $E_F$, the conductance of $WTe_2$/FGT is approximately $7.63 \times 10^{-6}$ S, but it increases dramatically to $7.80 \times 10^{-5}$ S when the energy level is shifted to -0.75 eV, reaching one-third of the conductance of FGT/FGT. This suggests that the $WTe_2$/FGT interface exhibits a substantial spin transparency ($T_{int}$). Although $T_{int}$ is lower than that in the heterostructures with chemical-binding interface interaction (e.g. Pt/FGT, see Section SV of the Supplemental Material [61] for details), the combination of $T_{int}$ with the large spin Hall angle of $WTe_2$ ($\theta_{SH} = 4.0$) [70] leads to the enhanced SOT efficiency observed experimentally [33]. Moreover, we predict that by applying an appropriate gate voltage, highly efficient spin transfer can be achieved between $WTe_2$ and FGT flacks.

Before concluding, we note that the energy difference between the bridge and hollow configurations of the $WTe_2$/FGT heterostructure is minimal, indicating the potential coexistence of both phases under experimental conditions. As shown in Fig. 8, a comparison of the energy band structures of the bridged and hollow (Fig. 5) configurations reveals that their shapes are very similar. The bridge configuration exhibits nearly the same polarization-dependent hybridization differences as that of hollow configuration. In addition, the transmission spectrum shows comparable differences in magnitude at the Fermi energy level, suggesting that both configurations, which can coexist experimentally, hold similar high interfacial spin injection properties.

## Ⅳ. CONCLUSION

In summary, based on the DFT calculations, we reveal the significant p-orbital hybridization between the interlayer Te atoms in the $WTe_2$/FGT vdW heterostructure. Such orbital hybridization significantly reduces the interface electronic potential barrier and modifies the spin-polarized electronic structure of FGT, resulting in the formation of an effective spin-polarized transport channel between $WTe_2$ and FGT. The NEGF calculations further demonstrate a large spin-polarized transmission, yielding a spin conductivity of $2.74\times10^4$ S/m for spin-up electrons across the $WTe_2$/FGT layers. This confirms the unusually high cross-plane spin transparency at the vdW interface. Given the large spin Hall angle of $WTe_2$, this finding naturally explains the remarkably high SOT efficiency observed in experiments. We also predict that higher SOT efficiency can be achieved through the application of gate voltage. This study provides an efficient approach to exploring and optimizing vdW heterostructures with high cross-plane spin transparency, which is particularly critical for designing high-performance all-vdW spintronic devices.

**Acknowledgments**

This work was supported by the Natural Science Foundation of China (No. 12474237), Science Fund for Distinguished Young Scholars of Shaanxi Province (No. 2024JC-JCQN-09), and Natural Science Foundation of Shaanxi Province (Grant No. 2023-JC-QN-0768).

## Reference

[1] L. Liu, C.-F. Pai, Y. Li, H. W. Tseng, D. C. Ralph, and R. A. Buhrman, Spin-torque switching with the giant spin Hall effect of tantalum, Science 336, 555 (2012).

[2] I. M. Miron, K. Garello, G. Gaudin, P.-J. Zermatten, M. V. Costache, S. Auffret, S. Bandiera, B. Rodmacq, A. Schuhl, and P. Gambardella, Perpendicular Switching of a Single Ferromagnetic Layer Induced by In-Plane Current Injection, Nature (London) 476, 189 (2011).

[3] J. Ryu, S. Lee, K. J. Lee, and B. G. Park, Current-induced spin-orbit torques for spintronic applications, Adv. Mater. 32, 1907148 (2020).

[4] Y. Fan, P. Upadhyaya, X. Kou, M. Lang, S. Takei, Z. Wang, J. Tang, L. He, L.-T. Chang, M. Montazeri, G. Yu, W. Jiang, T. Nie, R. N. Schwartz, Y. Tserkovnyak, and K. L. Wang, Magnetization switching through giant spin-orbit torque in a magnetically doped topological insulator heterostructure, Nat. Mater. 13, 699 (2014).

[5] H. Kurebayashi, J. Sinova, D. Fang, A. C. Irvine, T. D. Skinner, J. Wunderlich, V. Novák, R. P. Campion, B. L. Gallagher, E. K. Vehstedt, L. P. Zârbo, K. Výborný, A. J. Ferguson, and T. Jungwirth, An antidamping spin-orbit torque originating from the Berry curvature, Nat. Nanotechnol. 9, 211 (2014).

[6] P.-H. Lin, B.-Y. Yang, M.-H. Tsai, P.-C. Chen, K.-F. Huang, H.-H. Lin, and C.-H. Lai, Manipulating exchange bias by spin-orbit torque, Nat. Mater. 18, 335 (2019).

[7] C. K. Safeer, E. Jué, A. Lopez, L. Buda-Prejbeanu, S. Auffret, S. Pizzini, O. Boulle, I. M. Miron, and G. Gaudin, Spin-orbit torque magnetization switching controlled by geometry, Nat. Nanotechnol. 11, 143 (2016).

[8] S. Fukami, C. Zhang, S. DuttaGupta, A. Kurenkov, and H. Ohno, Magnetization Switching by Spin-Orbit Torque in an Antiferromagnet-Ferromagnet Bilayer System, Nat. Mater. 15, 535 (2016).

[9] F. Hellman, A. Hoffman, Y. Tserkovnyak, G. S. D. Beach, E. E. Fullerton, C. Leighton, A. H. MacDonald, D. C. Ralph, D. A. Arena, et al., Interface-induced phenomena in magnetism, Rev. Mod. Phys. 89, 025006 (2017).

[10] C. Zhang, S. Fukami, H. Sato, F. Matsukura, and H. Ohno, Spin-orbit torque induced magnetization switching in nano-scale Ta/CoFeB/MgO, Appl. Phys. Lett. 107, 012401 (2015).

[11] X. Qiu, K. Narayanapillai, Y. Wu, P. Deorani, D.-H. Yang, W.-S. Noh, J.-H. Park, K.-J. Lee, H.-W. Lee, and H. Yang, Spin-orbit torque engineering via oxygen manipulation, Nat. Nanotechnol. 10, 333 (2015).

[12] C. Safranski, E. A. Montoya, and I. N. Krivorotov, Spin–orbit torque driven by a planar Hall current, Nat. Nanotechnol. 14, 27 (2019).

[13] J. W. Lee, Y.-W. Oh, S.-Y. Park, A. I. Figueroa, G. van der Laan, G. Go, K.-J. Lee, and B.-G. Park, Enhanced spin-orbit torque by engineering Pt resistivity in $Pt/Co/AlO_x$ structures, Phys. Rev. B 96, 064405 (2017).

[14] J. Han, A. Richardella, S. A. Siddiqui, J. Finley, N. Samarth, and L. Liu, Room-Temperature Spin-Orbit Torque Switching Induced by a Topological Insulator, Phys. Rev. Lett. 119, 077702 (2017).

[15] N. H. D. Khang, Y. Ueda, and P. N. Hai, A conductive topological insulator with large spin Hall effect for ultralow power spin–orbit torque switching, Nat. Mater. 17, 808 (2018).

[16] M. DC, R. Grassi, J.-Y. Chen, M. Jamali, D. R. Hickey, D. Zhang, Z. Zhao, H. Li, P. Quarterman, Y. Lv, M. Li, A. Manchon, K. A. Mkhoyan, T. Low, and J.-P. Wang, Room-temperature high spin-orbit torque due to quantum confinement in sputtered $Bi_xSe_{(1-x)}$ films, Nat. Mater. 17, 800 (2018).

[17] D. MacNeill, G. M. Stiehl, M. H. D. Guimarães, N. D. Reynolds, R. A. Buhrman, and D. C. Ralph, Thickness dependence of spin-orbit torques generated by $WTe_2$, Phys. Rev. B 96, 054450 (2017).

[18] D. MacNeill, G. M. Stiehl, M. H. D. Guimaraes, R. A. Buhrman, J. Park, and D. C. Ralph, Control of spin–orbit torques through crystal symmetry in $WTe_2$/Ferromagnet bilayers, Nat. Phys. 13, 300 (2017).

[19] P. Li, W. Wu, Y. Wen, C. Zhang, J. Zhang, S. Zhang, Z. Yu, S. A. Yang, A. Manchon, and X. X. Zhang, Spin-momentum locking and spin-orbit torques in magnetic nano-heterojunctions composed of Weyl semimetal $WTe_2$, Nat. Commun. 9, 3990 (2018).

[20] B. Zhao, B. Karpiak, D. Khokhriakov, A. Johansson, A. M. Hoque, X. Xu, Y. Jiang, I. Mertig, and S. P. Dash, Unconventional charge–spin conversion in Weyl-semimetal $WTe_2$, Adv. Mater. 32, 2000818 (2020).

[21] C. Gong, L. Li, Z. Li, H. Ji, A. Stern, Y. Xia, T. Cao, W. Bao, C. Wang, Y. Wang et al., Discovery of intrinsic ferromagnetism in two-dimensional van der Waals crystals, Nature (London) 546, 265 (2017).

[22] B. Huang, G. Clark, E. Navarro-Moratalla, D. R. Klein, R. Cheng, K. L. Seyler, D. Zhong, E. Schmidgall, M. A. McGuire, D. H. Cobden et al., Layer-dependent ferromagnetism in a van der Waals crystal down to the monolayer limit, Nature (London) 546, 270 (2017).

[23] S. Jiang, L. Li, Z. Wang, K. F. Mak, and J. Shan, Controlling magnetism in 2D $CrI_3$ by electrostatic doping, Nat. Nanotechnol. 13, 549 (2018).

[24] Z. Wang, I. Gutiérrez-Lezama, N. Ubrig, M. Kroner, M. Gibertini, T. Taniguchi, K. Watanabe, A. Imamoğlu, E. Giannini, and A. F. Morpurgo, Very large tunneling magnetoresistance in layered magnetic semiconductor $CrI_3$, Nat. Commun. 9, 2516 (2018).

[25] G. T. Lin, X. Luo, F. C. Chen, J. Yan, J. J. Gao, Y. Sun, W. Tong, P. Tong, W. J. Lu, Z. G. Sheng, W. H. Song, X. B. Zhu, and Y. P. Sun, Critical behavior of two-dimensional intrinsically ferromagnetic semiconductor $CrI_3$, Appl. Phys. Lett. 112, 072405 (2018).

[26] Y. Sun, R. C. Xiao, G. T. Lin, R. R. Zhang, L. S. Ling, Z. W. Ma, X. Luo, W. J. Lu, Y. P. Sun, and Z. G. Sheng, Effects of hydrostatic pressure on spin-lattice coupling in two-dimensional ferromagnetic $Cr_2Ge_2Te_6$, Appl. Phys. Lett. 112, 072409 (2018).

[27] J. Zeisner, A. Alfonsov, S. Selter, S. Aswartham, M. P. Ghimire, M. Richter, J. van den Brink, B. Buchner, and V. Kataev, Magnetic anisotropy and spin-polarized two-dimensional electron gas in the van der Waals ferromagnet $Cr_2Ge_2Te_6$, Phys. Rev. B 99, 165109 (2019).

[28] M. Suzuki, B. Gao, K. Koshiishi, S. Nakata, K. Hagiwara, C. Lin, Y. X. Wan, H. Kumigashira, K. Ono, S. Kang, S. Kang, J. Yu, M. Kobayashi, S. W. Cheong, and A. Fujimori, Coulomb-interaction effect on the two-dimensional electronic structure of the van der Waals ferromagnet $Cr_2Ge_2Te_6$, Phys. Rev. B 99, 161401(R) (2019).

[29] K. Kim, J. Seo, E. Lee, K. T. Ko, B. S. Kim, B. G. Jang, J. M. Ok, J. Lee, Y. J. Jo, W. Kang, J. H. Shim, C. Kim, H. W. Yeom, B. Il Min, B. J. Yang, and J. S. Kim, Large anomalous Hall current induced by topological nodal lines in a ferromagnetic van der Waals semimetal, Nat. Mater. 17, 794 (2018).

[30] Z. Fei, B. Huang, P. Malinowski, W. Wang, T. Song, J. Sanchez, W. Yao, D. Xiao, X. Zhu, A. F. May, W. Wu, D. H. Cobden, J. H. Chu, and X. Xu, Two-dimensional itinerant ferromagnetism in atomically thin $Fe_3GeTe_2$, Nat. Mater. 17, 778 (2018).

[31] Y. Deng, Y. Yu, Y. Song, J. Zhang, N. Z. Wang, Z. Sun, Y. Yi, Y. Z. Wu, S. Wu, J. Zhu, J. Wang, X. H. Chen, and Y. Zhang, Gate-tunable room-temperature ferromagnetism in two-dimensional $Fe_3GeTe_2$, Nature (London) 563, 94 (2018).

[32] H.-J. Deiseroth, K. Aleksandrov, C. Reiner, L. Kienle, and R. K. Kremer, $Fe_3GeTe_2$ and $Ni_3GeTe_2$ – two new layered transition - metal compounds: crystal structures, HRTEM investigations, and magnetic and electrical properties, Eur. J. Inorg. Chem. 2006, 1561 (2006).

[33] I. Shin, W. J. Cho, E. S. An, S. Park, H. W. Jeong, S. Jang, W. J. Baek, S. Y. Park, D. H. Yang, J. H. Seo, G. Y. Kim, M. N. Ali, S. Y. Choi, H. W. Lee, J. S. Kim, S. D. Kim, and G. H. Lee, Spin-orbit torque switching in an all-van der Waals heterostructure, Adv. Mater.

34, 2101730 (2022).

[34] S. Hori, K. Ueda, T. Kida, M. Hagiwara, and J. Matsuno, Spin–orbit torque generation in bilayers composed of CoFeB and epitaxial $SrIrO_3$ grown on an orthorhombic $DyScO_3$ substrate, Appl. Phys. Lett. 121, 022402 (2022).

[35] Y. Zheng, T. Wang, X. Su, Y. Chen, Y. Wang, H. Lv, S. Cardoso, D. Yang, and J. Cao, Enhancement of spin-orbit torques in $Ta/Co_{20}Fe_{60}B_{20}/MgO$ structures induced by annealing, AIP Advances. 7, 075305 (2017).

[36] S. Shi, S. Liang, Z. Zhu, K. Cai, S. D. Pollard, Y. Wang, J. Wang, Q. Wang, P. He, J. Yu, G. Eda, G. Liang, and H. Yang, All-electric magnetization switching and Dzyaloshinskii–Moriya interaction in $WTe_2$/ferromagnet heterostructures, Nat. Nanotechnol. 14, 945 (2019).

[37] X. Wang, J. Tang, X. Xia, C. He, J. Zhang, Y. Liu, C. Wan, C. Fang, C. Guo, W. Yang, Y. Guang, X. Zhang, H. Xu, J. Wei, M. Liao, X. Lu, J. Feng, X. Li, Y. Peng, H. Wei1, R. Yang, D. Shi, X. Zhang, Z. Han, Z. Zhang, G. Zhang, G. Yu, and X. Han, Highly Efficient Spin–Orbit Torque and Switching of Layered Ferromagnet $Fe_3GeTe_2$, Sci. Adv. 5, eaaw8904 (2019).

[38] X. Li, P. Li, V. D. H. Hou, M. DC, C.-H. Nien, F. Xue, D. Yi, C. Bi, C.-M. Lee, S.-J. Lin, W. Tsai, Y. Suzuki, and S. X. Wang, Large and robust charge-to-spin conversion in sputtered conductive $WTe_x$ with disorder, Matter 4, 1639 (2021).

[39] G. Kresse and J. Furthmüller, Efficient iterative schemes for ab initio total-energy calculations using a plane-wave basis set, Phys. Rev. B 54, 11169 (1996).

[40] P. E. Blöchl, Projector augmented-wave method, Phys. Rev. B 50, 17953 (1994).

[41] G. Kresse and J. Hafner, Ab initio molecular dynamics for liquid metals, Phys. Rev. B 47, 558 (1993).

[42] J. P. Perdew, K. Burke, and M. Ernzerhof, Generalized Gradient Approximation Made Simple, Phys. Rev. Lett. 77, 3865 (1996).

[43] S. Grimme, Semiempirical GGA-type density functional constructed with a long-range dispersion correction, J. Comput. Chem. 27, 1787 (2006).

[44] H. J. Monkhorst and J. D. Pack, Special points for Brillouin-zone integrations, Phys. Rev. B 13, 5188 (1976).

[45] Z. X. Shen, X. Y. Bo, K. Cao, X. G. Wan, and L. X. He, Magnetic ground state and electron-doping tuning of Curie temperature in $Fe_3GeTe_2$: First-principles studies, Phys. Rev. B 103, 085102 (2021).

[46] L. Tang, Y. Zhou, C. Kang, B. Wen, J. Tian, and W. Zhang, Enhancing magnetocrystalline

anisotropy through interface coupling in a 2D ferromagnetic $Fe_3GeTe_2/VI_3$ heterostructure, Appl. Phys. Lett. 124, 012403 (2024).

[47] H. Wang, Y. Liu, P. Wu, W. Hou, Y. Jiang, X. Li, C. Pandey, D. Chen, Q. Yang, H. Wang, D. Wei, N. Lei, W. Kang, L. Wen, T. Nie, W. Zhao, and K. L. Wang, Above room-temperature ferromagnetism in wafer-scale two-dimensional van der waals $Fe_3GeTe_2$ tailored by a topological insulator, ACS Nano 14, 10045 (2020).

[48] J. Yang, R. Quhe, S. Liu, Y. Peng, X. Sun, L. Zha, B. Wu, B. Shi, C. Yang, J. Shi, G. Tian, C. Wang, J. Lu, and J. Yang, Gate-tunable high magnetoresistance in monolayer $Fe_3GeTe_2$ spin valves, Phys. Chem. Chem. Phys. 22, 25730 (2020).

[49] M. Q. Dong, Z. X. Song and Z. X. Guo, High-order anisotropic magnetoresistance in two-dimensional magnetic monolayers: First-principles study based on $Fe_3GeTe_2$ and $CrTe_2$, Phys. Rev. B 111, 174447 (2025).

[50] E. J. Sie, C. M. Nyby, C. D. Pemmaraju, S. J. Park, X. Shen, J. Yang, M. C. Hoffmann, B. K. Ofori-Okai, R. Li, A. H. Reid, S. Weathersby, E. Mannebach, N. Finney, D. Rhodes, D. Chenet, A. Antony, L. Balicas, J. Hone, T. P. Devereaux, T. F. Heinz, X. Wang and A. M. Lindenberg, An ultrafast symmetry switch in a Weyl semimetal, Nature (London) 565, 61 (2019).

[51] H.-J. Deiseroth, K. Aleksandrov, C. Reiner, L. Kienle, and R. K. Kremer, $Fe_3GeTe_2$ and $Ni_3GeTe_2$—Two New Layered Transition-Metal Compounds: Crystal Structures, HRTEM Investigations, and Magnetic and Electrical Properties, Eur. J. Inorg. Chem. 2006, 1561 (2006).

[52] M. Brandbyge, J.-L. Mozos, P. Ordejón, J. Taylor, and K. Stokbro, Density-functional method for nonequilibrium electron transport, Phys. Rev. B 65, 165401 (2002).

[53] J. S. Huang, P. Li, X. X. Ren, and Z.-X. Guo, Promising properties of a sub-5-nm monolayer $MoSi_2N_4$ transistor, Phys. Rev. Appl. 16, 044022 (2021).

[54] X. Zhang, B. Liu, J. S. Huang, X. W. Cao, Y. Z. Zhang, and Z.-X. Guo, Nonvolatile spin field effect transistor based on $VSi_2N_4/Sc_2CO_2$ heterostructure, Phys. Rev. B 109, 205105 (2024).

[55] J. P. Perdew, K. Burke, and M. Ernzerhof, Generalized Gradient Approximation Made Simple, Phys. Rev. Lett. 77, 3865 (1996).

[56] F. Ye, J. Lee, J. Hu, Z. Mao, J. Wei, and P. X.-L. Feng, Environmental instability and degradation of single- and few-layer $WTe_2$ nanosheets in ambient conditions, Small 12, 5802 (2016).

[57] Y. Deng, Y. Yu, Y. Song, J. Zhang, N. Z. Wang, Z. Sun, Y. Yi, Y. Z. Wu, S. Wu, J. Zhu et

al., Gate-tunable room-temperature ferromagnetism in two-dimensional $Fe_3GeTe_2$, Nature 563, 94 (2018).

[58] Z. Fei, B. Huang, P. Malinowski, W. Wang, T. Song, J. Sanchez, W. Yao, D. Xiao, X. Zhu, A. F. May et al., Two-dimensional itinerant ferromagnetism in atomically thin $Fe_3GeTe_2$, Nat. Mater. 17, 778 (2018).

[59] X. Wang, J. Tang, X. Xia, C. He, J. Zhang, Y. Liu, C. Wan, C. Fang, C. Guo, W. Yang et al., Current-driven magnetization switching in a van der Waals ferromagnet $Fe_3GeTe_2$, Sci. Adv. 5, eaaw8904 (2019).

[60] X. Kong, G. D. Nguyen, J. Lee, C. Lee, S. Calder, A. F. May, Z. Gai, A. P. Li, L. Liang, and T. Berlijn, Magnetic ground state and electron-doping tuning of Curie temperature in $Fe_3GeTe_2$: First-principles studies, Phys. Rev. Materials 4, 094403 (2020).

[61] See supplemental material at http://link.aps.org/supplemental/10.1103/xxx for Orbital-projected band structures of $WTe_2$ Three configurations of $WTe_2$/FGT heterostructure, Energy comparison of collinear vs con-collinear configuration, Phonon spectrum calculations for $WTe_2$/FGT heterostructure, The interfacial spin properties of the Pt/FGT heterostructure, Comparison with previous results. The supplemental material also contains Refs. [64-71].

[62] I. H. Kao, R. Muzzio, H. Zhang, M. Zhu, J. Gobbo, S. Yuan, D. Weber, R. Rao, J. Li, J. H. Edgar, J. E. Goldberger, J. Yan, D. G. Mandrus, J. Hwang, R. Cheng, J. Katoch, and S. Singh, Deterministic switching of a perpendicularly polarized magnet using unconventional spin-orbit torques in $WTe_2$, Nat. Mater. 21, 1029 (2022).

[63] Y. Wu, S. Zhang, J. Zhang, W. Wang, Y. L. Zhu, J. Hu, K. Wong, C. Fang, C. Wan, X. Han et al., Néel-type skyrmion in $WTe_2/Fe_3GeTe_2$ van der Waals heterostructure, Nat. Commun. 11, 3860 (2020).

[64] B. Marfoua, J. Hong, Highly efficient spin-orbit torque generation in bilayer $WTe_2/Fe_3GaTe_2$ heterostructure, Mater. Today Phys. 42, 101378 (2024).

[65] Y. Wang, P. Deorani, X. Qiu, J. H. Kwon, and H. Yang, Determination of intrinsic spin Hall angle in Pt, Appl. Phys. Lett. 105, 152412 (2014).

[66] G. V. Pushkarev, D. I. Badrtdinov, I. A. Iakovlev, V. V. Mazurenko, and A. N. Rudenko, An effective spin model on the honeycomb lattice for the description of magnetic properties in two-dimensional $Fe_3GeTe_2$, J. Magn. Magn. Mater. 588, 171456 (2023).

[67] A. M. Ruiz, D. L. Esteras, D. Lopez-Alcala, and J. J. Baldov, On the origin of the above room-temperature magnetism in the 2D van der Waals ferromagnet $Fe_3GeTe_2$, Nano Lett. 24, 7886 (2024).

[68] D. Li, S. Haldar, T. Drevelow, and S. Heinze, Tuning the magnetic interactions in van der Waals $Fe_3GeTe_2$ heterostructures: A comparative study of ab initio methods, Phys. Rev. B 107, 104428 (2023).

[69] Q. Zhao, Y. Zhu, H. Zhang, B. Jiang, Y. Wang, T. Xie, K. Lou, C. Xia, H. Yang, and C. Bi, Proximity-induced interfacial room-temperature ferromagnetism in semiconducting $Fe_3GeTe_2$, ACS Applied Materials & Interfaces 15, 46520 (2023).

[70] K. V. Zakharchenko, J. H. Los, M. I. Katsnelson, and A. Fasolino, Atomistic simulations of structural and thermodynamic properties of bilayer graphene, Phys. Rev. B 81, 235439 (2010).

[71] H. Rydberg, M. Dion, N. Jacobson, E. Schroder, P. Hyldgaard, S. I. Simak, D. C. Langreth, and B. I. Lundqvist, Van der Waals Density Functional for Layered Structures, Phys. Rev. Lett. 91, 126402 (2003).

[72] F. Lou, X. Y. Li, J. Y. Ji, H. Y. Yu, J. Feng, X. G. Gong, and H. J. Xiang, PASP: Property analysis and simulation package for materials, J. Chem. Phys. 154, 114103 (2021).

[73] P. Li, X. S. Zhou, and Z. X. Guo, Intriguing magnetoelectric effect in two-dimensional ferromagnetic/perovskite oxide ferroelectric heterostructure, npj Comput. Mater. 8, 20 (2022).

[74] T. V. Vu, T. P. Dao, M. Idrees, H. V. Phuc, N. N. Hieu, N. T. Binh, H. B. Dinh, B. Amin, and C. V. Nguyen, Effects of different surface functionalization on the electronic properties and contact types of graphene/functionalized-GeC van der Waals heterostructures, Phys. Chem. Chem. Phys. 22, 7952 (2020).

[75] X. Zhang, L. Feng, S. Zhong, Y. Ye, H. Pan, P. Liu, X. Zheng, H. Li, M. Qu, and X. Wang, Schottky barrier heights and mechanism of charge transfer at metal-$Bi_2OS_2$ interfaces, Sci. China Mater. 66, 811 (2023).

[76] L. Cao, X. Deng, Z. Tang, R. Tan and Y. S. Ang, Designing CMOS compatible efficient ohmic contacts to $WSi_2N_4$ via surface-engineered $Mo_2B$ monolayer electrodes, J. Mater. Chem. C. 12, 648 (2024).

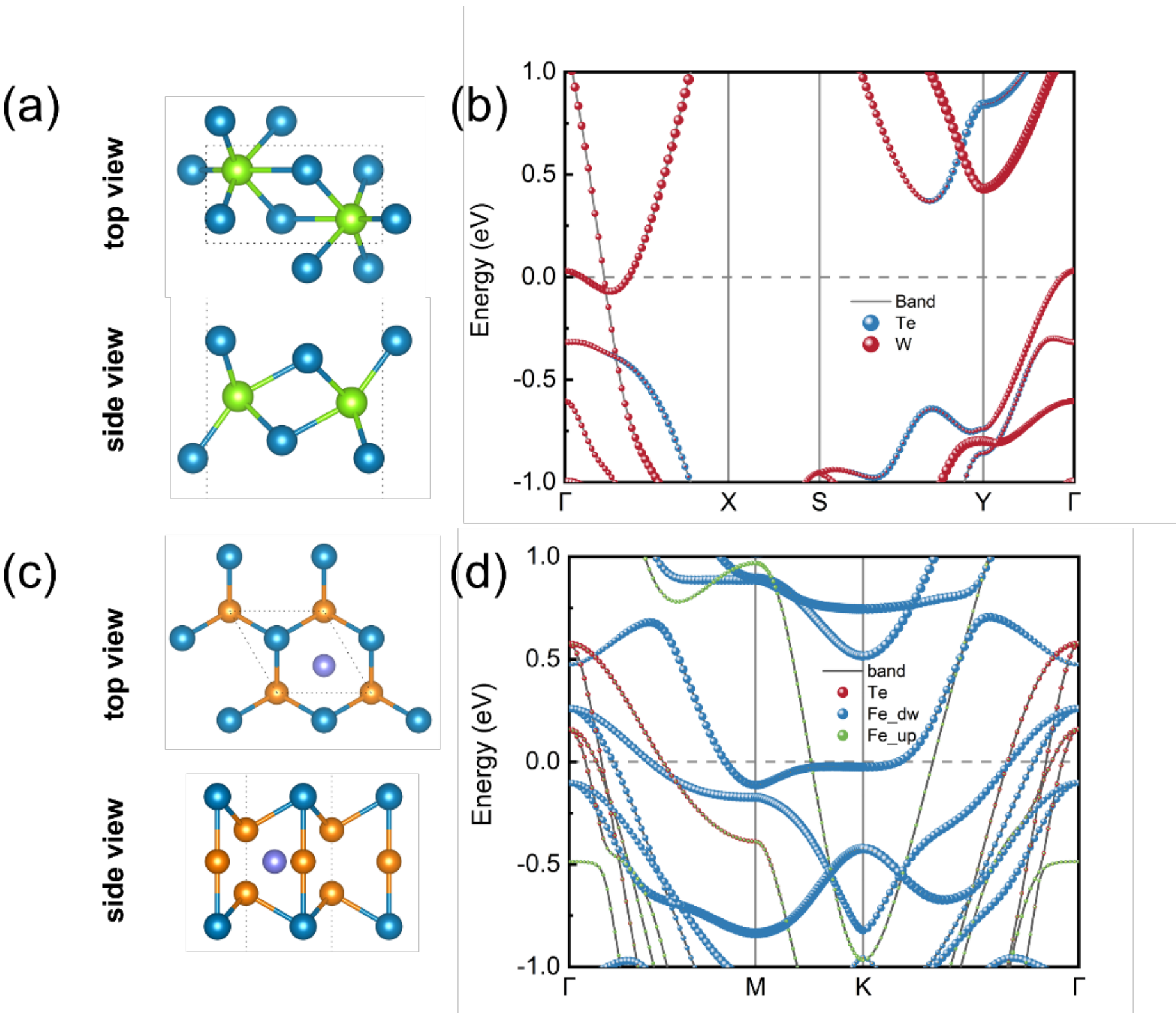

FIG. 1. The atomic structures and elemental-projected energy band structures of monolayer $WTe_2$ (a-b) and monolayer FGT (c-d), respectively, based on DFT calculations.

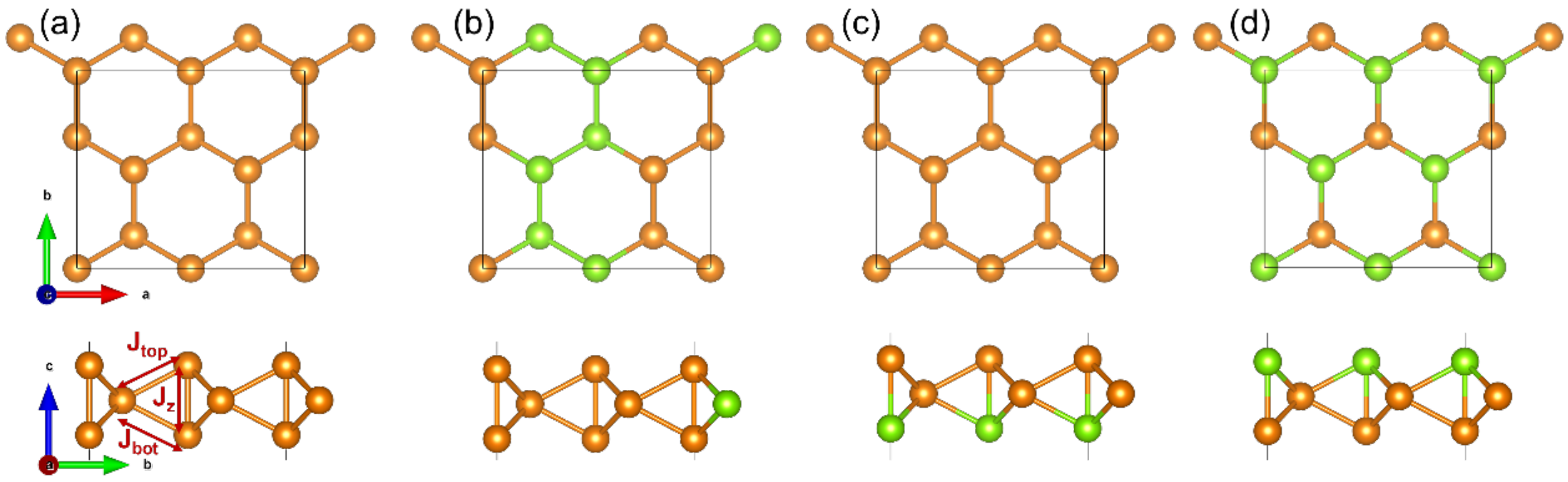

FIG. 2. Schematics of typical magnetic configurations of FGT (only Fe atoms are shown), including one FM (a) and three AFMs: AFM1 (b), AFM2 (c), AFM3 (d). The FGT is composed of three layers of Fe atoms, with $Fe^{2+}$ ions in the middle layer and $Fe^{3+}$ ions in the top and bottom layers. The green and orange colors denote magnetic moments oriented in the +z and -z directions, respectively.

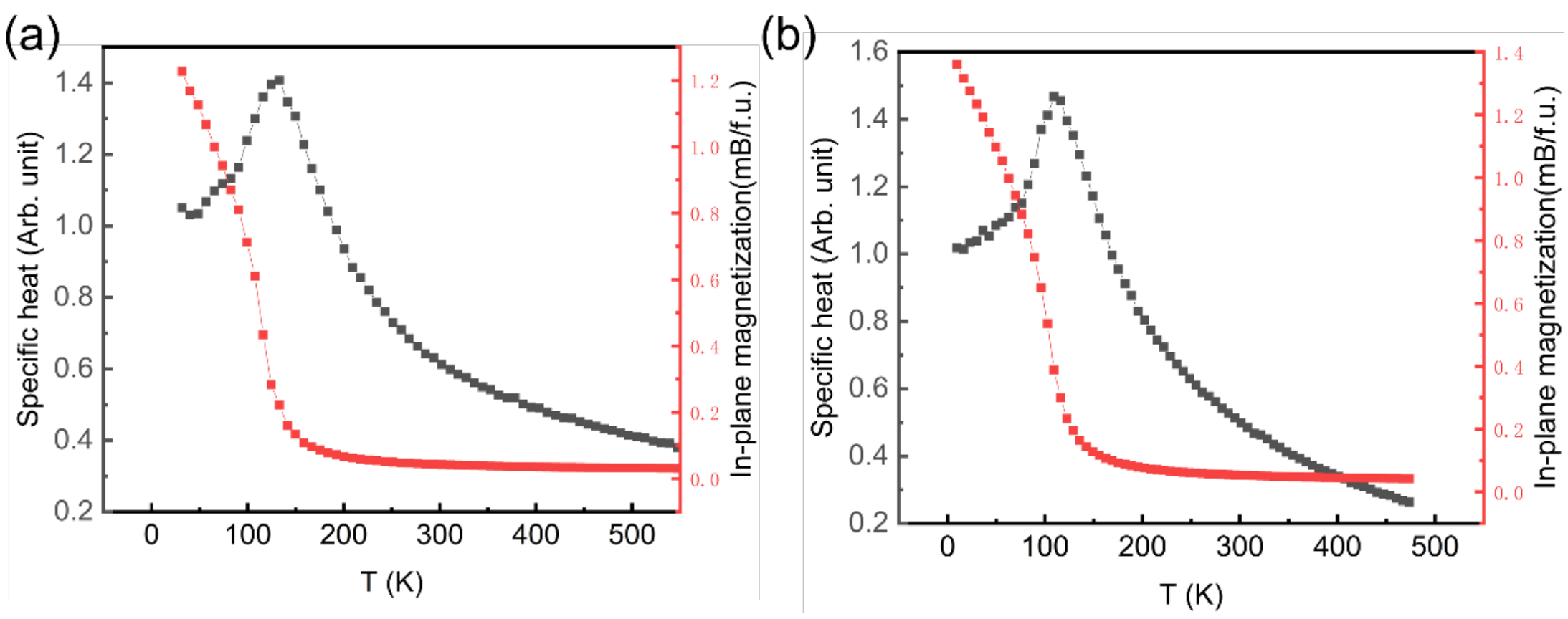

FIG. 3. The calculated specific heat (black line) and magnetization (red line) as a function of temperature for (a) the $WTe_2$/FGT heterostructure and (b) the freestanding FGT layer.

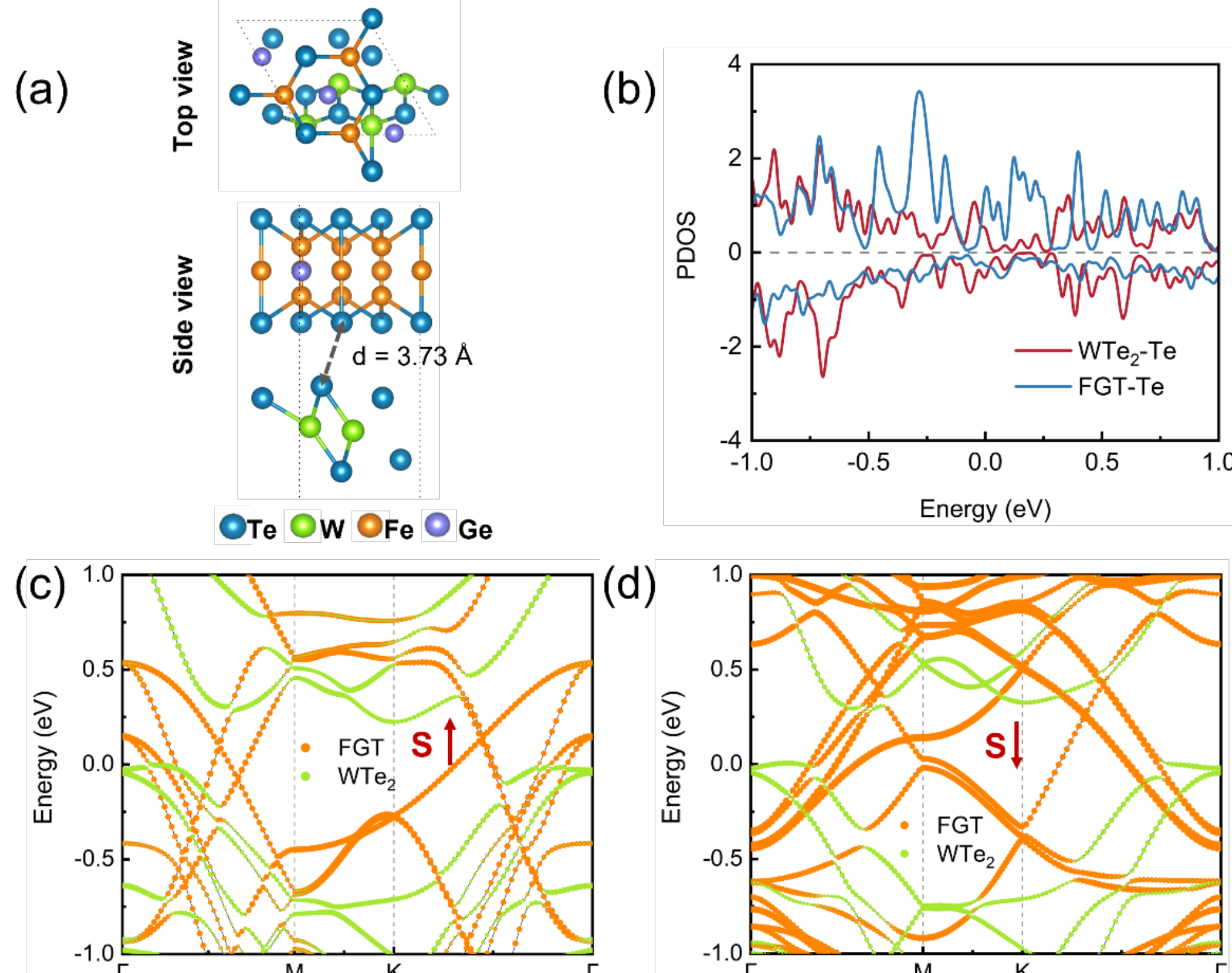

FIG. 4. (a) Crystal structure of the $WTe_2$/FGT heterostructure. Blue, green, orange, and purple spheres represent Te, W, Fe, and Ge atoms, respectively. (b) Partial density of states (PDOS) of Te atoms at the interface of the $WTe_2$/FGT heterostructure. Positive and negative values on the vertical axis correspond to spin-up and spin-down PDOS, respectively. (c) and (d) Spin-up and spin-down projected electronic band structures of the $WTe_2$/FGT heterostructure. Orange and green dots indicate the orbital contributions from FGT and $WTe_2$, respectively. Note that the projections are limited to the interfacial Te atoms. Red arrows denote spin polarization directions.

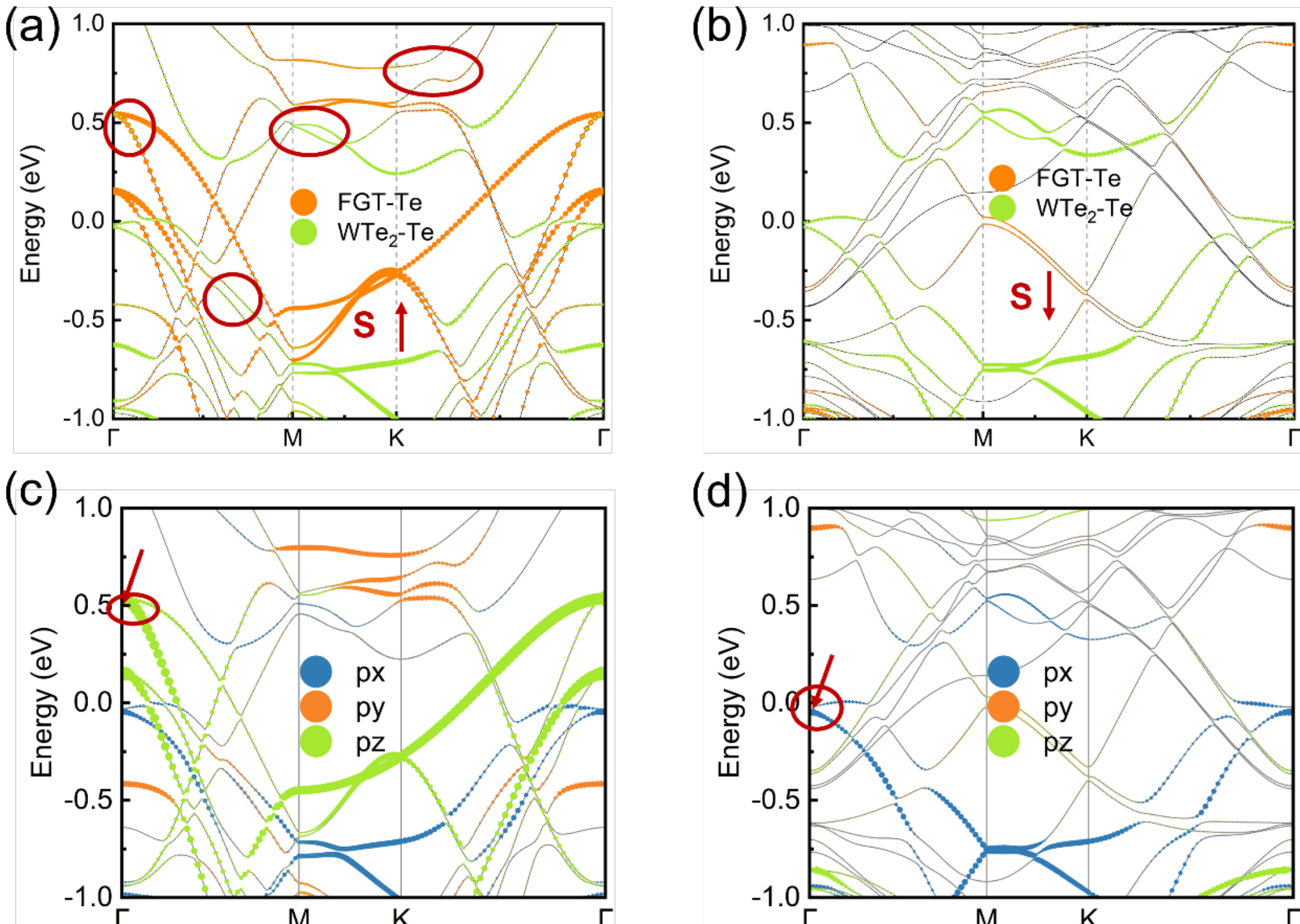

FIG. 5. (a) and (b) are the spin-up and spin-down projected electronic band structures on Te atoms at the interface of the $WTe_2$/FGT heterostructure, respectively. Orange and green points represent the contributions of FGT and $WTe_2$, respectively. Red circles in (a) highlight regions of band hybridization. (c) and (d) present the $p_m$-orbital (m=x,y,z) projected band structures for spin-up and spin-down electrons, respectively. The magnitude of projection is calculated by using the product of $W_{pm1} \times W_{pm2}$, where $W_{pm1}$ and $W_{pm2}$ are the weights of $p_m$ orbitals of Te atoms of FGT and $WTe_2$ at the interface, respectively. Bule ($p_x$), orange ($p_y$) and green ($p_z$) dots represent $p_x$-$p_x$, $p_y$-$p_y$, $p_z$-$p_z$ orbital hybridizations at the interface, respectively. The red circles in (c) and (d) indicate the selected energy band at Γ point for the wave-function visualization, shown in Figs. 4 (a) and 4(b), respectively.

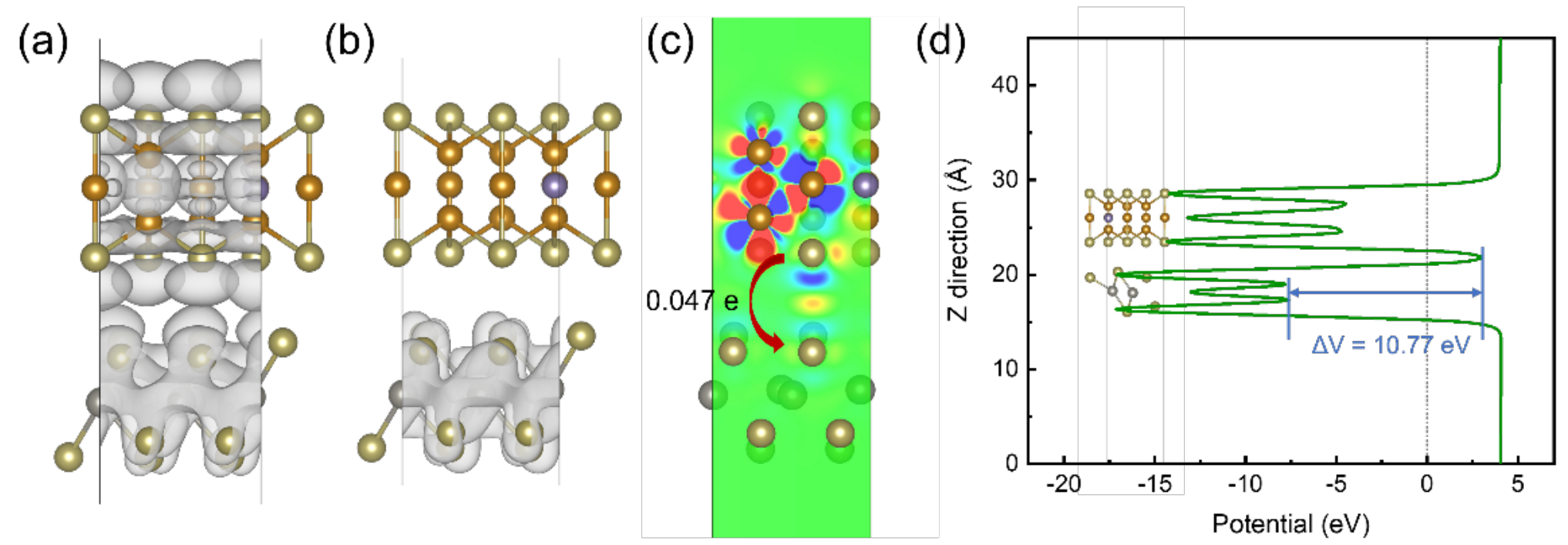

FIG. 6. (a-b) Visualization of squared wave-functions of the spin-up (a) and spin-down (b) electronic states nearby the Fermi level (as indicated in Figs. 5(c) and 5(d)), respectively. An isosurface value of $5\times10^{-12}$ e/Bohr$^3$ is used. (c) Differential charge density in (010) surface. (d) Electrostatic potential of $WTe_2$/FGT heterostructure along Z direction.

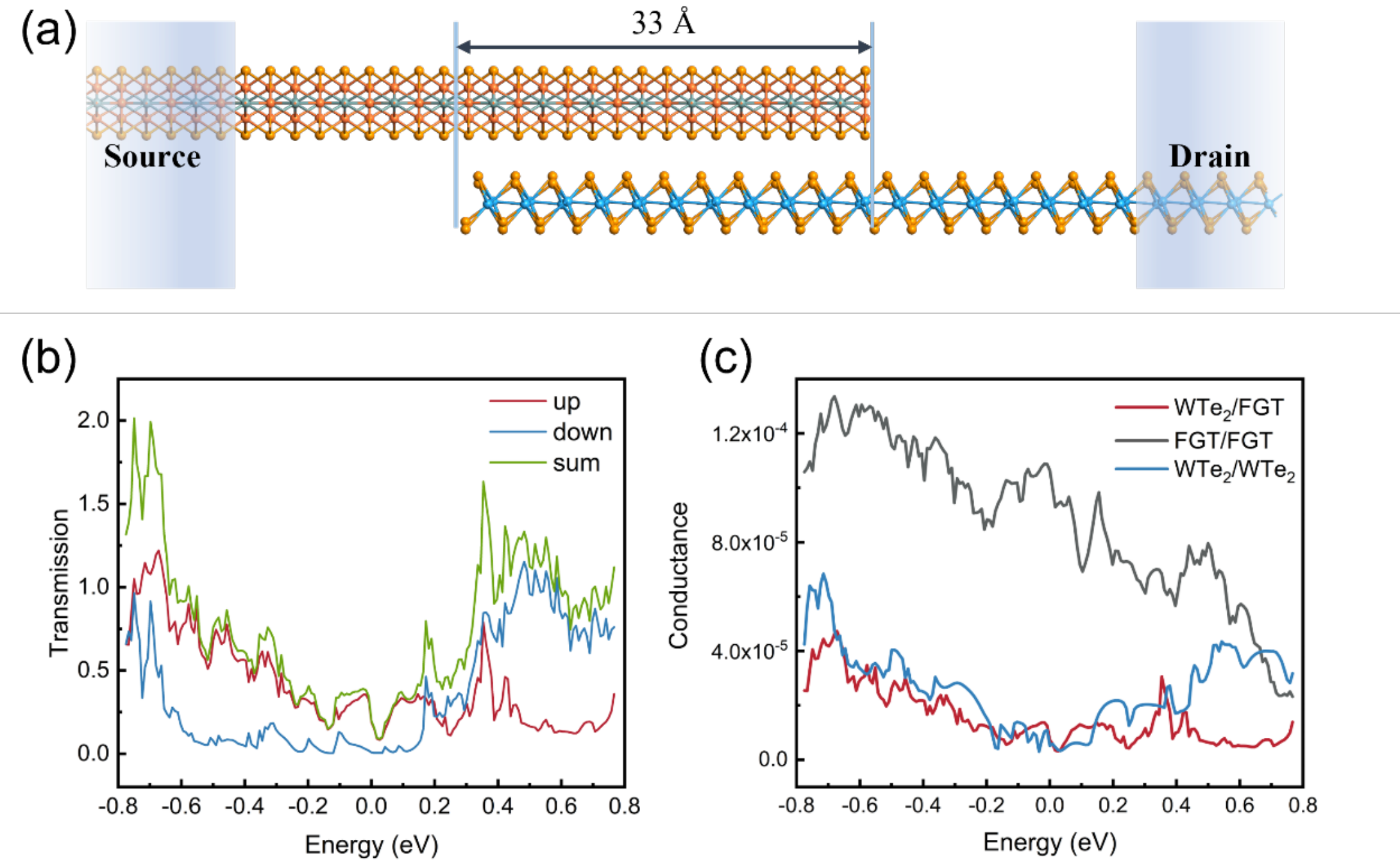

FIG. 7. (a) Device structure of $WTe_2$/FGT heterostructure. The overlapping region in the middle is about 33 Å. (b) The spin-dependent transmission spectrum of $WTe_2$/FGT device at zero bias. (c) Device spin-up conductance. Red, black, and blue lines represent $WTe_2$/FGT device, FGT/FGT device and $WTe_2$/$WTe_2$ device, respectively.

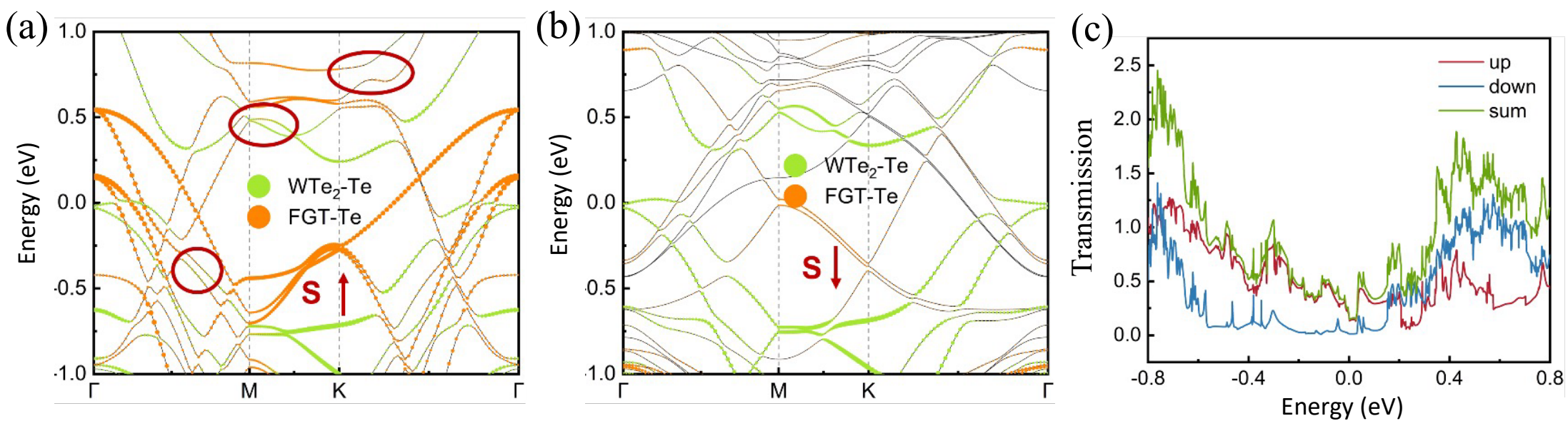

FIG. 8. (a-b) Spin-up and spin-down projected electronic band structures of the $WTe_2$/FGT heterostructure in the bridge stacking configuration, respectively. Note that the projections are performed to only the Te atoms at the interface. Green and orange points represent the contributions of $WTe_2$ and FGT, respectively. The red arrows represent the direction of spin polarization. Red circles indicate band hybridization. (c) The spin-dependent transmission spectrum of $WTe_2$/FGT device in the bridge stacking configuration at zero bias.

Table 1. Total energy of monolayer FGT and $WTe_2$/FGT under different magnetic configurations.

| | FM | AFM1 | AFM2 | AFM3 |
|---|---|---|---|---|
| Monolayer FGT | -107.4366 eV | -106.1731 eV | -106.6849 eV | -106.6850 eV |
| $WTe_2$/FGT | -662.7995 eV | -660.9799 eV | -658.9938 eV | -658.6127 eV |